\documentclass[pdflatex,sn-mathphys-num]{sn-jnl}% Math and Physical Sciences Numbered Reference Style
\usepackage{graphicx}%
\usepackage{multirow}%
\usepackage{amsmath,amssymb,amsfonts}%
\usepackage{amsthm}%
\usepackage{mathrsfs}%
\usepackage[title]{appendix}%
\usepackage{xcolor}%
\usepackage{textcomp}%
\usepackage{manyfoot}%
\usepackage{booktabs}%
\usepackage{algorithm}%
\usepackage{algorithmicx}%
\usepackage{algpseudocode}%
\usepackage{listings}%
\def\re{\mathop{\rm Re}}
\def\im{\mathop{\rm Im}}

\theoremstyle{thmstyleone}%
\newtheorem{theorem}{Theorem}%  meant for continuous numbers
\theoremstyle{thmstyletwo}%

\theoremstyle{thmstylethree}%

\begin{document}
%%%%%%%%%%%%%%%%%%%%%%%%%%%%%%%%%%%%%%%%%%%%%%%%%%%%%%%%%%%%%%%%%%%%%%%
\title[Shrinking of the Szeg\H{o} curve]{The Schwarz function and the shrinking of the Szeg\H{o} curve:
                                                              electrostatic, hydrodynamic, and random matrix models}
%%%%%%%%%%%%%%%%%%%%%%%%%%%%%%%%%%%%%%%%%%%%%%%%%%%%%%%%%%%%%%%%%%%%%%%
%%=============================================================%%
%% GivenName	-> \fnm{Joergen W.}
%% Particle	-> \spfx{van der} -> surname prefix
%% FamilyName	-> \sur{Ploeg}
%% Suffix	-> \sfx{IV}
%% \author*[1,2]{\fnm{Joergen W.} \spfx{van der} \sur{Ploeg} 
%%  \sfx{IV}}\email{iauthor@gmail.com}
%%=============================================================%%

\author[1]{\fnm{Gabriel} \sur{\'Alvarez}}\email{galvarez@ucm.es}
\equalcont{These authors contributed equally to this work.}

\author[1]{\fnm{Luis} \sur{Mart\'{\i}nez Alonso}}\email{luism@ucm.es}
\equalcont{These authors contributed equally to this work.}

\author*[2]{\fnm{Elena} \sur{Medina}}\email{elena.medina@uca.es}
\equalcont{These authors contributed equally to this work.}

\affil[1]{\orgdiv{Departamento de F\'{\i}sica Te\'orica}, \orgname{Universidad Complutense de Madrid}, \orgaddress{\street{Plaza de Ciencias 1}, \city{Madrid}, \postcode{28040}, \country{Spain}}}

\affil*[2]{\orgdiv{Departamento de Matem\'aticas}, \orgname{Universidad de C\'adiz}, \orgaddress{\street{Campus Universitario R\'{\i}o San Pedro}, \city{C\'adiz}, \postcode{11510}, \country{Spain}}}

%%==================================%%
%% Sample for unstructured abstract %%
%%==================================%%

\abstract{We study the deformation of the classical Szeg\H{o} curve $\gamma_0$ given by
$\gamma_t  = \{ z\in\mathbb{C}: |z\, e^{1-z}| = e^{-t}, |z|\leq 1\}$, $t\geq 0$ from three different
viewpoints: an electrostatic equilibrium problem, the dual hydrodynamic model, and a
random matrix model. The common framework underlying these models is the
asymptotic distribution of zeros of the scaled varying Laguerre polynomials $L^{(\alpha_n)}_n(n z)$
in the critical regime where $\lim_{n\to\infty}\alpha_n/n=-1$, for which the limiting zero distribution
is supported on $\gamma_t$, where the deformation parameter $t$ encodes the exponential rate
at which the sequence $\alpha_n$ approximates the set of negative integers. We show that the Schwarz functions
of these curves can be written in terms of the Lambert $W$ function, and that in this formulation the $S$-property
of Stahl and Gonchar and Rachmanov can be explictly written as the Schwarz reflection symmetry.
We also discuss a conformal map of the interior of the curves $\gamma_t$ onto the disks $D(0,e^{-t})$
and the harmonic moments of the curves.}


%%\pacs[JEL Classification]{D8, H51}

%%\pacs[MSC Classification]{35A01, 65L10, 65L12, 65L20, 65L70}

%%%%%%%%%%%%%%%%%%%%%%%%%%%%%%%%%%%%%%%%%%%%%%%%%%%%%%%%%%%%%%%%%%%%%%%
\maketitle
%%%%%%%%%%%%%%%%%%%%%%%%%%%%%%%%%%%%%%%%%%%%%%%%%%%%%%%%%%%%%%%%%%%%%%%
%%%%%%%%%%%%%%%%%%%%%%%%%%%%%%%%%%%%%%%%%%%%%%%%%%%%%%%%%%%%
%% INTRODUCTION %%%%%%%%%%%%%%%%%%%%%%%%%%%%%%%%%%%%%%%%%%%%%%%%
%%%%%%%%%%%%%%%%%%%%%%%%%%%%%%%%%%%%%%%%%%%%%%%%%%%%%%%%%%%%
\section{Introduction}
%%%%%%%%%%%%%%%%%%%%%%%%%%%%%%%%%%%%%%%%%%%%%%%%%%%%%%%%%%%%
In a classical~1924 work Szeg\H{o}~\cite{SZ24} proved that the zeros of the rescaled partial sums of the
exponential series, i.e., the zeros of the polynomials
\begin{equation}
	p_n(nz) = \sum_{k=0}^n \frac{(n z)^k}{k!},
\end{equation}
accumulate as $n\to\infty$ on the curve in the complex $z$-plane given by
\begin{equation}
     \label{szcurve}
    |z e^{1-z}|=1, \quad |z|\leq 1,
\end{equation}
which is known as the Szeg\H{o} curve.
The partial sums of the exponential series are, up to a sign, the Laguerre polynomials $L^{(-n-1)}_n(z)$,
\begin{equation}
    p_n(z) = (-1) ^n L^{(-n-1)}_n(z),
\end{equation}
where
\begin{equation}
    \label{lagp}
    L^{(\alpha)}_n(z) = \sum_{k=0}^n \binom{n+\alpha}{n-k} \frac{(-z)^k}{k!},
\end{equation}
and Szeg\H{o}'s result is now understood as a special case within the general theory of the asymptotic
distribution of zeros of the scaled varying Laguerre polynomials $L^{(\alpha_n)}_n(nz)$,
where the sequence of real numbers $\alpha_n$ is such that
\begin{equation}
   \lim_{n\to\infty} \frac{\alpha_n}{n} = A,
\end{equation}
with $A$ a finite constant~\cite{MA01,KU01,KU04,DI11}. Szeg\H{o}'s result is the special case
$\alpha_n = -n-1$, which leads to $A=-1$.

In general, for a given a sequence of monic polynomials $\{S_n(z)\}_{n=1}^\infty$ where
\begin{equation}
	\label{fep}
	S_n(z) = \prod_{i=1}^n (z-z_i^{(n)}),
\end{equation}
there is an associated sequence of zero counting measures 
\begin{equation}
	\label{me1}
	\nu_n = \sum_{i=1}^n\delta(z-z_i^{(n)}),
\end{equation}
and the asymptotic distribution of zeros is described by the limit
of the sequence of normalized measures~\cite{ST10,MAMF07,MA11}
\begin{equation}
    \label{mun}
    \mu_n = \frac{1}{n}\nu_n
\end{equation}
as $n\rightarrow\infty$ in the sense of the weak-* topology~\cite{ST10,BI71}. 
Since the set of unit measures with uniformly bounded supports is compact in the weak-* topology,
if all the zeros $z_i^{(n)}$ belong to the same compact set $K\subset\mathbb{C}$,
then there exists a unit measure $\mu$ supported on $K$ and a subsequence $\Delta\subset\mathbb{N}$
such that $\mu_n\overset{*}{\rightarrow}\mu$ for $n\in\Delta$.

This setting for sequences of general orthogonal polynomials is discussed in detail in Chapter~2 of
the classical monograph~\cite{ST10} and, indeed, if $\alpha < -1$ is not an integer, the
Laguerre polynomials $L^{(\alpha)}_n(z)$ satisfy a nonhermitian orthogonality relation
\begin{equation}
	\int_{\widetilde{\Gamma}}  L^{(\alpha)}_n(z) z^k z^\alpha e^{-z} dz=0,\quad k=0,1,\ldots,n-1,
\end{equation}
for Hankel-type paths  $\widetilde{\Gamma}$  on $\mathbb{C} \setminus [0,+\infty)$ with
endpoints $+\infty\mp i \epsilon$ ($\epsilon>0$), where the integral is understood by analytic continuation
along $\widetilde{\Gamma}$ of any branch of the integrand.

The deformations of the Szeg\H{o} curve that are the subject of this paper appear as supports
of the limit of the zero counting measures of scaled varying Laguerre polynomials
$L^{(\alpha_n)}_n(nz)$ for certain sequences $\alpha_n$. In fact, the asymptotic distribution of zeros has been
thoroughly analyzed for all values of $A\in\mathbb{R}$, and a key result, established in~\cite{KU04,DI11}, is that for $-1\leq A\leq  0$
the support of the asymptotic zero density depends not only on the value of $A$ but also on an additional parameter  $t$,
which quantifies the exponential rate at which the sequence $\alpha_n$ approximates the set of negative integers.
In particular, the results of D\'{\i}az-Mendoza and Orive~\cite{DI11} on the asymptotic distribution of zeros of $L^{(\alpha_n)}_n(n z)$
for the  critical case $A=-1$ can be summarized as follows:
\begin{theorem}
\label{theo}
Let $\mathbb{S}_n=\{-1,-2,\ldots,-n\}$. If $A=-1$ and the limit
 \begin{equation}
     \label{tdef}
     e^{-t} = \lim_{n\to\infty}\big(\mathrm{dist}(\alpha_n, \mathbb{S}_n)\big)^{1/n},
 \end{equation}
exists with $0\leq t\leq +\infty$, then the normalized asymptotic zero density $\rho_t(z)$ and its support $\gamma_t$ are given by: 
\begin{description}
\item[(a)] For $0\leq t <+\infty$
     \begin{equation}
         \label{ze40}
         d\mu_t(z) = \rho_t(z)|{ d} z|=\frac{1}{2\pi i}\frac{1-z}{z}{ d} z,
    \end{equation}
   with
    \begin{equation}
        \label{zeo0dup}
        \gamma_t  = \{ z\in\mathbb{C}: |z\, e^{1-z}| = e^{-t}, \quad |z|\leq 1\}.
 \end{equation}
\item[(b)] 
For $t=+\infty$, $\rho_{\infty}(z)=\delta(z)$, with $\gamma_{\infty}=\{0\}$.
\end{description}
\end{theorem}

For example, if for any $c\geq 0$ we set
\begin{equation}
	\label{e2}
	\alpha_n=-n-c,\quad  n=1,2,\ldots,
\end{equation}
then $A=-1$ and $\mathrm{dist}(\alpha_n, \mathbb{S}_n) = c$, so that
\begin{equation}\label{e13}
	\lim_{n\to\infty}\big(\mathrm{dist}(\alpha_n, \mathbb{S}_n)\big)^{1/n}
	=
	\left\{\begin{array}{c}0,\quad \mbox{if $c=0$} \\1,\quad \mbox{if $c\neq 0$}  \end{array}\right.. 
\end{equation}
Therefore for $c=0$  we are in case (b), while for $c>0$ we are in case (a) with $t=0$. 
The classical result by Szeg\H{o} is recovered for $c=1$. We note that the Szeg\H{o} curve~(\ref{szcurve})
and the family of curves~(\ref{zeo0dup}) also appear in Refs.~\cite{LE17,LE23} as ``generalized Szeg\H{o} curves,''
namely curves along which the zeros of orthogonal polynomials with respect to certain complex measures accumulate.

%%%%%%%%%%%%%%%%%%%%%%%%%%%%%%%%%%%%%%%%%%%%%%%%%%%%%%%%%%%%
\begin{figure}[h]
\centering
\includegraphics[width=0.75\textwidth]{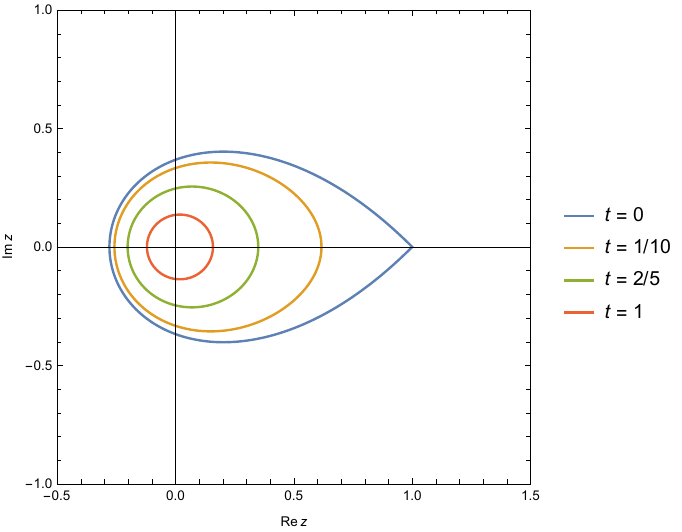}
\caption{Curves $\gamma_t$ given by $|z e^{1-z}| = e^{-t}$, $|z|\leq 1$, for $t=0, 1/10, 2/5$ and $1$.
              The curve $\gamma_0$ is the Szeg\H{o} curve.}
\label{fig1}
\end{figure}
%%%%%%%%%%%%%%%%%%%%%%%%%%%%%%%%%%%%%%%%%%%%%%%%%%%%%%%%%%%%

In Fig.~\ref{fig1} we show the curves $\gamma_t$ for $t=0$ (the Szeg\H{o} curve~(\ref{szcurve})),
$t=1/10, 2/5$ and $1$. It is apparent from this figure (and we will prove later) that as $t\rightarrow +\infty$
the curves $\gamma_t$ shrink to the origin $z=0$ with curvatures tending to a constant value. 

The main purpose of this work is to analyze the $t$-parametrized family of curves $\gamma_t$
from three different viewpoints: an electrostatic model, the dual hydrodynamic model, and a random matrix model.

Results of Stahl~\cite{ST86a,ST86b} and Gonchar and Rakhmanov~\cite{GO89} developed
further in Ref.~\cite{MAMF07} prove that the asymptotic normalized zero density
$\rho(z)$ with support $\gamma\subset\Gamma$ can be interpreted as an equilibrium electrostatic
problem in the presence of an external field, where the total electrostatic energy of the line conductor
$\gamma$  is given by
\begin{equation}
    \label{min}
    E = \int_{\gamma}(\re z -A\log |z|) \rho(z) | d z|
           -
           \int_{\gamma}\int_{\gamma} \log |z-z'| \rho(z) \rho(z') | d z|| d z'|.
\end{equation}
The first term is the energy due to the interaction with a uniform field and with a point charge $A/2$ at the origin,
i.e., with the external field, and the second term is the self-energy of the line conductor,
which we denote by $E_\mathrm{se}$:
\begin{equation}
    E_\mathrm{se} = -\int_{\gamma}\int_{\gamma} \log |z-z'| \rho(z) \rho(z') | d z|| d z'|.
\end{equation}
Note that we follow the conventions of Saff and~Totik~\cite{ST2E}, which differ by a factor of 2 with
respect to the physically more accurate convention of Ref.~\cite{MA01}, but leads to simpler equations.
(Incidentally, there has been recent interest in equilibrium electrostatic models associated with generalized Laguerre
polynomials defined by more general scalar products~\cite{HU14,DO24,DI24}, as well as in models related
to multiple orthogonal polynomials~\cite{MF23}.) This electrostatic equilibrium problem in the presence
of an external field can be conveniently formulated in terms of the total electrostatic potential,
 \begin{equation}
    \label{uz}
    U(z) = \re z  - A \log|z| -2 \int_{\gamma} \log |z-z'| \rho(z') |dz'|,
\end{equation}
by imposing  the conditions,
\begin{eqnarray}
     \label{u0a}
     U(z) & = & u_0,\quad z\in\gamma, \\
     U(z) & \geq & u_0,\quad  z\in\Gamma,
\end{eqnarray}
for a certain constant $u_0$, where $\Gamma$ is the Hankel-type path that satisfies the $S$-property,
\begin{equation}
    \label{eq:sprop}
    \frac{\partial U}{\partial \mathbf{n}_+}(z) = \frac{\partial U}{\partial \mathbf{n}_-}(z), \quad z\in \Gamma,
\end{equation}
and $\mathbf{n}_+ = - \mathbf{n}_-$ are the normals to $\Gamma$.
(To be precise, the results of Stahl and of Gonchar and Rakhmanov cannot be directly applied to
our problem because the complement of the support $\gamma$ is not connected, but D\'{\i}az-Mendoza and Orive~\cite{DI11}
showed that these results still hold.) Note that using equation~(\ref{u0a}), the total electrostatic energy
of the line conductor can be also written as
\begin{equation}
    \label{eu0mse}
    E = u_0-{E}_\mathrm{se}.
\end{equation} 

When we specialize to our case of interest,
namely the case where $A=-1$ and the limit~(\ref{tdef}) exists, the support of the equilibrium
measures are indeed the curves $\gamma_t$ for finite $t>0 $. We will show that the associated
potential functions, electric fields, and characteristic energies can be computed explicitly.
In particular, the parameter $t$ is related to the self-energy $E_\mathrm{se}$ of the line conductor
$\gamma_t$ supporting $\rho_t$ by
\begin{equation}
    \label{ce}
    E_\mathrm{se}=t+1.
\end{equation}
All these results have an immediate application to the dual hydrodynamical model.

Our third and last viewpoint starts with random matrix models with partition functions of
the form
\begin{equation}
	\label{mm}
	Z_n(g)
	=
	\int_{\Gamma\times\cdots\times\Gamma}
	\prod_{j<k}(z_j-z_k)^2
	\exp\left(-\frac{1}{g}\sum_{i=1}^n W(z_i)\right)
	\prod_{i=1}^n  d z_i,
\end{equation}
for sequences of parameters $g_n$ such that the limit
\begin{equation}
   \label{tH}
   \lim_{n\rightarrow\infty}ng_n=T
\end{equation}
exists ($T$ is known as the {}'t~Hooft parameter). We give arguments valid for any matrix model
such that $W'(z)$ is a rational function of the complex variable $z$ whose only singularities are
at most a finite number of simple poles $\mathcal{A}=\{a_1,\ldots,a_m\}$.
We also assume that the integration path $\Gamma$ lies in the domain of analyticity of $W(z)$
and that the integral is convergent. The particular case
\begin{equation}
    \label{eq:pmm}
    W(z) = z + \log z
\end{equation}
is known as the Penner matrix model, and the associated polynomials whose zeros
are the saddle points of the partition function turn out again to be proportional
to the scaled varying Laguerre polynomials $L^{(\alpha_n)}_n(z/g_n)$ where $\alpha_n = - 1 - 1/g_n$.
The critical case corresponds to $T=1$.

The layout of the paper is the following. In section~\ref{sec:cmef} and by restricting ourselves
to the critical case ($A=-1$ in the electrostatic interpretation, $T=1$ in the Penner matrix
model interpretation), we give a brief derivation of the appearance of the deformations
of the Szeg\H{o}  curves $\gamma_t$ as supports of the critical measures. This derivation is entirely
independent of the theory of Laguerre polynomials, and therefore does not provide
the interpretation of the parameter $t$ in Theorem~\ref{theo}.
In section~\ref{sec:sch} we discuss the Schwarz
function of the curves $\gamma_t$ and several magnitudes of interest that can be computed
in terms of it. In section~\ref{sec:elec} we present the results pertaining to the electrostatic
model, including the corresponding results for the dual hydrodynamic model.
Finally, in section~\ref{sec:mm} we discuss in a rather general setting the
saddle points for a random matrix model and the Schwinger-Dyson equation,
and then particularize for the Penner matrix model~(\ref{eq:pmm}) and
more precisely to the {}'t~Hooft limit of the critical Penner matrix model.
%%%%%%%%%%%%%%%%%%%%%%%%%%%%%%%%%%%%%%%%%%%%%%%%%%%%%%%%%%%%
%% CRITICAL MEASURE %%%%%%%%%%%%%%%%%%%%%%%%%%%%%%%%%%%%%%%%%%%%%%%
%%%%%%%%%%%%%%%%%%%%%%%%%%%%%%%%%%%%%%%%%%%%%%%%%%%%%%%%%%%%
\section{Critical measure in the external field\label{sec:cmef}}
%%%%%%%%%%%%%%%%%%%%%%%%%%%%%%%%%%%%%%%%%%%%%%%%%%%%%%%%%%%%
In this section we restrict ourselves to the critical case ($A=-1$ in the electrostatic interpretation, $T=1$ in the Penner matrix
model interpretation) and find the equilibrium measures corresponding to the complex potential~(\ref{eq:pmm}) within the
framework of the formalism developed in Ref.~\cite{MA11}.

The total electrostatic potential~(\ref{uz}) takes the form
\begin{equation}
    U(z) =  \re z + \log|z| - 2\int_{\gamma}\log|z-z'|\rho(z')|dz'|,
\end{equation}
and the complex electrostatic potential on $\gamma$ can be written as
\begin{equation}
    \Omega(z) = W(z)-\left(g(z_+)+g(z_-)\right),\quad z\in\gamma,
\end{equation}
where
\begin{equation}
    g(z) = \int_{\gamma}\log(z-z')\rho(z')|dz'|.
\end{equation}
The $S$-property~(\ref{eq:sprop}) reads
\begin{equation}
   \Omega'(z) = 0,\quad z\in\gamma,
\end{equation}
or, equivalently,
\begin{equation}
    \label{eq:sprop2}
    W'(z)-\left(g'(z_+)+g'(z_-)\right) = 0,\quad z\in\gamma.
\end{equation}
Anticipating our treatment of matrix models in section~\ref{sec:mm},
we define a function
\begin{equation}
    \label{y}
    y(z) = W'(z)-2g'(z).
\end{equation}
Note that
\begin{equation}
    g'(z) = \int_{\gamma}\frac{1}{z-z'}\rho(z')|dz'|
\end{equation}
is analytic in $\mathbb{C}\setminus\gamma$, and that in terms of $y(z)$ the $S$-property~(\ref{eq:sprop2})
reads
\begin{equation}
    \label{yc}
    y(z_+)=-y(z_-),\quad z\in\gamma.
\end{equation}
Taking into account that $W'(z) = 1 + 1/z$, it turns out that the function
\begin{equation}
    R(z) = y(z)^2
\end{equation}
is analytic in $\mathbb{C}\setminus\{0\}$ and the isolated singularity as $z=0$ is a double pole in which the
coefficient of $1/z^2$ is 1. Liouville's theorem implies that $R(z)$ is a rational function with denominator $z^2$,
and since
\begin{equation}
    R(z)=\left(1+\frac{1}{z}-2\left(\frac{1}{z}+\mathcal{O}\left(\frac{1}{z^2}\right)\right)\right)^2,\quad \mbox{as} \quad
z\rightarrow\infty,
\end{equation}
we conclude that
\begin{equation}
    \label{eq:rcm}
    R(z) = \left(1-\frac{1}{z}\right)^2.
\end{equation}

The supports of the critical measures are the trajectories of the quadratic differential~\cite{MA11,DI11}
\begin{equation}
    -R(z)(dz)^2,
\end{equation}
or 
\begin{equation}
    \re\int_1^z\sqrt{R(z')}\,dz' 
    =
    \re\int_1^z \left( 1 - \frac{1}{z'}\right)\,dz'
    =
    t, \quad 0\leq t < \infty
\end{equation}
which is equivalent to
\begin{equation}
    \label{eq:zlz}
    \re(z - \log z) = t + 1,
\end{equation}
and therefore to equation~(\ref{zeo0dup}) for $\gamma_t$.
Equation~(\ref{eq:zlz}) can be rewritten as
\begin{equation}
    \re(z - \log z) = x_0 - \log x_0,\quad 0 < x_0 \le 1,
\end{equation}
where $x_0 - \log x_0 = t + 1$, which shows explicitly that 
the Szeg\H{o} curve $\gamma_0$ (with $x_0=1$) is a critical
trajectory of the quadratic differential stemming from the simple zero of $R(z)$
at $z=1$, while the remaining curves $\gamma_t$ are regular trajectories (ovals)
encircling the pole of $R(z)$ at $z=0$. The corresponding critical densities
are given by
\begin{equation}
    \rho_t(z)|dz| = \frac{1}{2\pi i}\sqrt{R(z)}dz=\frac{1}{2\pi i}\frac{1-z}{z}dz.
\end{equation}
Note that they are formally independent of $t$ and that Cauchy's theorem confirms
that they are normalized. The limiting case $t=\infty$ corresponds to the measure
$d\mu_\infty(z) = \rho_\infty(z)\, dz = \delta(z) \,dz$. We refer for details to Refs.~\cite{MA11,DI11},
where in addition it is shown that the measures $\mu_t(z)$ for $0\leq t<\infty$ are
the balayages (as defined, e.g., in Ref.~\cite{ST2E}) of $\mu_\infty(z)$ from the interior
of $\gamma_t$ onto $\gamma_t$ (Lemma~2.1 in Ref.~\cite{DI11}).
%%%%%%%%%%%%%%%%%%%%%%%%%%%%%%%%%%%%%%%%%%%%%%%%%%%%%%%%%%%%
%% SCHWARZ %%%%%%%%%%%%%%%%%%%%%%%%%%%%%%%%%%%%%%%%%%%%%%%%%%%%
%%%%%%%%%%%%%%%%%%%%%%%%%%%%%%%%%%%%%%%%%%%%%%%%%%%%%%%%%%%%
\section{The Schwarz function and the shrinking process\label{sec:sch}}
%%%%%%%%%%%%%%%%%%%%%%%%%%%%%%%%%%%%%%%%%%%%%%%%%%%%%%%%%%%%
%%%%%%%%%%%%%%%%%%%%%%%%%%%%%%%%%%%%%%%%%%%%%%%%%%%%%%%%%%%%
As we mentioned in the Introduction, the Schwarz function of the curves $\gamma_t$ can be explicitly written in terms
of the Lambert $W$ function~\cite{CO96,OL10}, which is implicitly defined by
\begin{equation}
    \label{dl}
    W(z) e^{W(z)} = z.
\end{equation}
On $[0,\infty)$ there is only one real solution, while on $(-1/e,0)$ there are two real solutions. The principal
branch, denoted by $\mathrm{Wp}$ in~\cite{OL10} but by $\mathrm{W}_0$ in this work,
is the solution that satisfies $W(x)\geq W(-1/e)$.
This branch is analytic in $\mathbb{C}\setminus (-\infty, -1/e]$, strictly increasing on $(-1/e,\infty)$,
maps the real interval $[-1/e,+\infty)$ onto $[-1,+\infty)$, and its Taylor series around the origin reads
\begin{equation}
    \label{tl}
    \mathrm{W}_0(z) = \sum_{n=1}^{\infty} (-1)^{n-1} \frac{n^{n-2}}{(n-1)!}z^n,\quad |z|<1/e.
\end{equation}
%%%%%%%%%%%%%%%%%%%%%%%%%%%%%%%%%%%%%%%%%%%%%%%%%%%%%%%%%%%%
\subsection{The Schwarz function of $\gamma_t$\label{sec:s}}
%%%%%%%%%%%%%%%%%%%%%%%%%%%%%%%%%%%%%%%%%%%%%%%%%%%%%%%%%%%%
Several interesting properties of the shrinking process can be easily derived in terms of the Schwarz function $S(z,t)$
of the curves $\gamma_t$, which are defined by the condition~\cite{DA74}
\begin{equation}
    \label{sf}
    \bar{z} = S(z,t) \quad\mbox{iff $z\in\gamma_t$}.
\end{equation}
From equation~(\ref{zeo0dup}) it follows that the Schwarz function of $\gamma_t$ can be written
in terms of the principal branch $\mathrm{W}_0$ of the Lambert $W$ function,
\begin{equation}
    \label{swla}
    S(z,t) = -\mathrm{W}_0 (\mathfrak{z}(z, t)),
\end{equation}
where
\begin{equation}
    \mathfrak{z}(z, t) = - e^{-2(t+1)} z^{-1} e^z.
\end{equation}

\begin{figure}[h]
\centering
\includegraphics[width=0.45\textwidth]{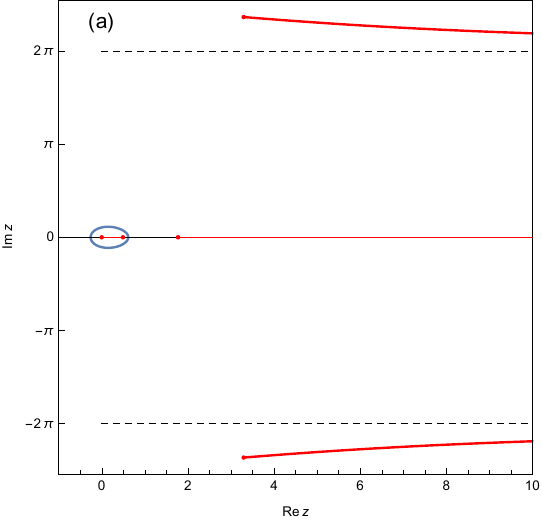}
\includegraphics[width=0.45\textwidth]{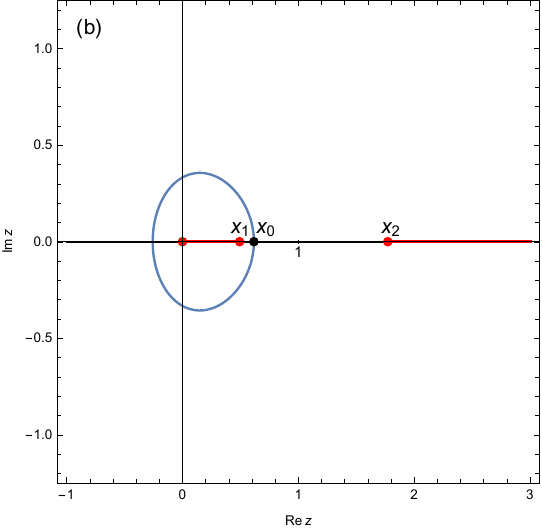}
\caption{(a) Domain of analyticity of the $S$ function for $\gamma_t$ with $t=1/10$ showing the two
branch cuts $[0,x_1]$ and $[x_2,\infty)$ on the nonnegative real axis and the first two
of the infinitely many branch cuts away from the real axis.
(b) Magnification of a neighborhood of the origin. The point labeled $x_0$ is the intersection of the
curve $\gamma_t$ with the positive real axis. Note that the Schwarz function is analytic in particular
on the annulus $x_1 < |z| < x_2$, which contains the unit circle $|z|=1$, and that the curve $\gamma_t$
is homotopic to the unit circle on $D(S(z,t))$.
\label{fig2}}
\end{figure}

In general, the Schwarz function $S(z)$ of an analytic curve is known to be analytic only in a strip-like neighborhood
of the curve, and for points $z$ close enough to the curve, the Schwarz reflection $z^*$ of $z$ across the curve is
defined by
\begin{equation}
	\label{la1}
	z^*=\overline{S(z)}.
\end{equation}
However, because of the explicit form~(\ref{swla}), the domain of analyticity of $S(z,t)$, which we will denote by $D(S(z,t))$,
can also be described very explicitly. The function $S(z,t)$ will be analytic in the complex $z = x+i y$ plane except for cuts
where
\begin{equation}
    \label{eq:fkz}
    \mathfrak{z}(z, t) = -\frac{e^{-2 t+x-2} }{x^2+y^2} \Big((x \cos y + y \sin y) - i (y \cos y-x \sin y)\Big)
\end{equation}    
is real and belongs to $(-\infty, -1/e]$. By setting $y=0$ in equation~(\ref{eq:fkz}), we find that this condition
yields two cuts on the nonnegative real axis: one from $z=0$ to the smallest solution $x_1<1$ of the equation
\begin{equation}
	\frac{e^x}{x} = e^{2t+1},
\end{equation}
and a second cut from the largest solution $x_2>1$ of this equation to $+\infty$. For $ y\neq 0$, $\mathfrak{z}(z, t)$
is also real when
\begin{equation}
	 y \cos y = x \sin y.
\end{equation}
The corresponding curves can be parametrized by $y$ as
\begin{equation}
	z = y \cot y + i y,
\end{equation}
and the cuts run from $y=2k\pi$ to the solution (if it exists) of
\begin{equation}
	\label{eq:bp}
	e^{y \cot y} \frac{\sin y}{y} = e^{2t+1},
\end{equation}
if $y>0$, or from the solution of equation~(\ref{eq:bp}) (if it exists) to $y=-2k\pi$ if $y<0$.
It is easy to see that equation~(\ref{eq:bp})
has exactly one solution on each $y$ interval $(2k\pi, (2k+1)\pi)$ with $k=1, 2, \ldots$, as well as the symmetric
solutions for negative $y$. This situation is illustrated in Fig.~\ref{fig2}~(a), where we show the curve
$\gamma_t$ for $t=1/10$, the two cuts on the nonnegative real axis, the cut corresponding to $k=1$
and its symmetric cut on the lower half-plane. Note, in particular, that these latter cuts are further away from the curve than
the cut $[x_2,\infty)$ on the real axis. Note also that the origin is a logarithmic branch point for $S(z,t)$ and
that the remaining branch points are algebraic of order 2~\cite{CO96}.

%%%%%%%%%%%%%%%%%%%%%%%%%%%%%%%%%%%%%%%%%%%%%%%%%%%%%%%%%%%%
%%%%%%%%%%%%%%%%%%%%%%%%%%%%%%%%%%%%%%%%%%%%%%%%%%%%%%%%%%%%

In Fig.~\ref{fig2}~(b) we show a magnification of a neighborhood of the origin. Note that the Schwarz function
is analytic in particular on the annulus $x_1 < |z| < x_2$, which contains the unit circle $|z|=1$, and that the curve
$\gamma_t$ is homotopic to the unit circle on $D(S(z,t))$. These facts allow us to obtain also expressions
for the corresponding harmonic moments of $\gamma_t$.  We denote by $D_t^+$ and $D_t^-$ the interior
and the exterior domains of the positively oriented curve $\gamma_t$, respectively.
The harmonic moments of the curves $\gamma_t$ are defined by
\begin{eqnarray}
   C_k(t) & = & -\frac{1}{ \pi} \int\!\!\!\int_{D_t^-} z^{-k}   dx dy, \quad k=1, 2, \ldots, \\
   C_{-k}(t) & = & \frac{1}{ \pi  } \int\!\!\!\int_{D_t^+} z^k  dx dy,\quad k=0,1, 2, \ldots.
\end{eqnarray}

Because of the defining property~(\ref{sf}), the harmonic moments can be written in terms of the
Schwarz function as
\begin{equation}
	\label{hm1}
 	C_k(t) = \frac{1}{ 2\pi  i} \oint_{\gamma_t}z^{-k}\,\bar{z}\,dz
	          = \frac{1}{ 2\pi  i} \oint_{\gamma_t}z^{-k}\,S(z,t) dz,  \quad k\in\mathbb{Z},
 \end{equation}
and because of the domain of analyticity of $S(z,t)$ and the fact that $\gamma_t$ is homotopic to the
unit circle in this domain, the Laurent expansion
 \begin{equation}
 	\label{lau}
 	S(z,t) = \sum_{k=-\infty}^{+\infty} C_k(t) z^{k-1},
 \end{equation}
can be obtained from the Taylor expansion of $\mathrm{W}_0(z)$ given in equation~(\ref{tl}),
which leads to
\begin{equation}
    \label{mmdup}
    C_k(t) = \sum_{n= {\rm max}(1,1-k)}^{\infty}\frac{n^{k+2n-2}}{(n+k-1)!\,n!} e^{-2(t+1) n}.	
\end{equation} 
Incidentally, the set of harmonic moments of $\gamma_t$ determines the electrostatic potential due
to a constant charge density filling  $D_t^+$.

Finally and for later reference, we mention two additional consequences of the explicit form~(\ref{swla}).
First, the relation between the partial derivatives
\begin{equation}
    \label{ld}
    \frac{\partial S}{\partial t} = \frac{2 z}{1-z} \frac{\partial S}{\partial z}.
\end{equation}
Second, equations~(\ref{la1}) and~(\ref{dl}) show that for the Schwarz reflection $z^*$ of $z$ on $\gamma_t$,
\begin{equation}
	\label{la3}
	z e^{-z} \overline{z^*} e^{-\overline{z^*}} = e^{-2(t+1)}.
\end{equation}
%%%%%%%%%%%%%%%%%%%%%%%%%%%%%%%%%%%%%%%%%%%%%%%%%%%%%%%%%%%%
\subsection{Conformal map and parametrization of $\gamma_t$\label{sec:cm}}
%%%%%%%%%%%%%%%%%%%%%%%%%%%%%%%%%%%%%%%%%%%%%%%%%%%%%%%%%%%%
There is a natural univalent conformal map $w(z)$ between $D_t^+$ and the open disk $|w|<e^{-t}$ given by
\begin{equation}
    w(z) = z e^{1-z}.
\end{equation}
The inverse of $w(z)$ can be expressed in terms of  the principal branch of the Lambert $W$ function,
\begin{equation}
	\label{z1}
  	z(w) = -\mathrm{W}_0(-w/e), \quad |w|<e^{-t}. 
 \end{equation}  
The map extends continuously to the boundary $|w|=e^{-t}$, which is mapped onto the boundary $\gamma_t$
of $D_t^+$. Note that the points of $\gamma_t$ satisfy the bound
\begin{equation}
	\label{in}
  	|z| \leq -\mathrm{W}_0(-e^{-(t+1)}),\quad z\in\gamma_t,
\end{equation}
with equality reached at $x_0=-\mathrm{W}_0(-e^{-(t+1)})$. Thus, the family of curves $\gamma_t$ can be parametrized by
\begin{equation}
	\label{z11}
  	z(t,\theta) = -\mathrm{W}_0(-e^{-(t+1)+i\theta}), \quad 0\leq t<+\infty, \; 0\leq \theta<2\pi. 
\end{equation} 
Since $z(t,\theta)$ depends on $t$ and $\theta$ through the combination $t-i\theta$,
\begin{equation}
    \label{lan2}
    \frac{\partial z}{\partial t}=i\, \frac{\partial z}{\partial \theta},
\end{equation}
which geometrically means that the points of $\gamma_t$  move with a normal inwards velocity. 
Moreover, since the Lambert function satisfies~\cite{CO96,OL10}
 \begin{equation}
 	\label{lan1}
 	\mathrm{W}_0'(z) = \frac{e^{-\mathrm{W}_0(z)}}{1+\mathrm{W}_0(z)},\quad z\in\mathbb{C}\setminus (-\infty, -1/e],
 \end{equation}
 we get that 
 \begin{equation}\label{lan2dup}
 \frac{\partial z}{\partial t}=i\, \frac{\partial z}{\partial \theta}=\frac{z}{z-1}.
 \end{equation}

\begin{figure}[h]
\centering
\includegraphics[width=0.75\textwidth]{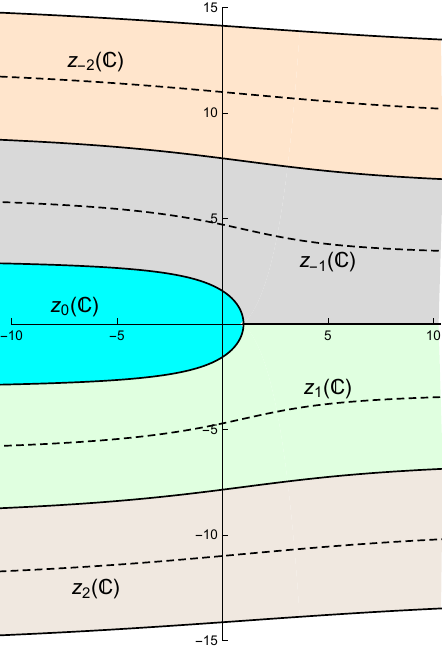}
\caption{Univalence regions bounded by the inverse images of the branch cuts of $\mathrm{W}_k(-w/e)$
             for $k=-2,-1,0,1,2$. The solid lines except the $[1,\infty)$ interval of the real axis, and
             the dashed lines form the \emph{Quadratrix of Hippias}.\label{fig3}}
\end{figure}

To extend the map $w(z)$ to a conformal one-to-one map on the whole $z$-plane we use
the Riemann surface  $\mathcal{R}_w$ of the inverse function $z(w)$.  The maximal regions
of the $z$-plane in which $w(z)$ is univalent can be labelled by an integer $k$, and the corresponding
inverse functions of $w(z)$ are 
\begin{equation}
	\label{zk}
  	z_k(w) = -\mathrm{W}_k(-w/e), \quad k\in\mathbb{Z}, 
 \end{equation}
where $\mathrm{W}_k$ denote the branches of the Lambert $W$ function. Hence, the boundary curves
that maximally partition the $z$-plane into univalence  regions are the inverse images of the branch cuts
of $\mathrm{W}_k(-w/e)$. These branch cuts $\omega_k$ are the real intervals $1\leq w<+\infty$ for $k=0$,
$ 0\leq w<+\infty$ for $k\neq 0,\pm1$ and a double cut along the intervals $1\leq w<+\infty$ and $0\leq w<+\infty$
for $k=\pm 1$~\cite{CO96}. The images of these cuts in the $z$-plane  form a subset of the
\emph{Quadratrix of Hippias}~\cite{CO96}. We illustrate these univalence regions in Fig.~\ref{fig3}.

%%%%%%%%%%%%%%%%%%%%%%%%%%%%%%%%%%%%%%%%%%%%%%%%%%%%%%%%%%%%
%%%%%%%%%%%%%%%%%%%%%%%%%%%%%%%%%%%%%%%%%%%%%%%%%%%%%%%%%%%%

%%%%%%%%%%%%%%%%%%%%%%%%%%%%%%%%%%%%%%%%%%%%%%%%%%%%%%%%%%%%
\subsection{Curvature of $\gamma_t$}
%%%%%%%%%%%%%%%%%%%%%%%%%%%%%%%%%%%%%%%%%%%%%%%%%%%%%%%%%%%%
The parametrization~(\ref{z11}) of the curves $\gamma_t$ permits a straightforward computation of their
unsigned curvatures $\kappa(t,\theta)$ in terms of the Schwarz $S$ function~\cite{DA74},
\begin{equation}
    \kappa(t,\theta) = \frac{1}{2} \big| S_{zz}( z(t,\theta), t) \big|,
\end{equation}
which, using equation~(\ref{lan1}), leads to
\begin{equation}
    \kappa(t,\theta) = \left|
                                           \frac{1+ \re(\mathrm{W}_0(\zeta)(2+\mathrm{W}_0(\zeta)))
                                           }{
                                           (1+\mathrm{W}_0(\bar{\zeta}))^3 \mathrm{W}_0(\zeta)}
                                \right|,
\end{equation}
where
\begin{equation}
    \zeta = - e^{-(t+1)}e^{i\theta}.
\end{equation}
Equation~(\ref{tl}) shows that
\begin{equation}
     \kappa(t,\theta) \sim e^{t+1} - e^{-(t+1)} \left(1 - \frac{3}{2} \cos 2\theta \right),
     \quad
     t \to \infty,
\end{equation}
i.e., that the shrinking process leads to curves with asymptotically constant curvature.
This behavior is illustrated in Fig.~\ref{fig1}, where the innermost curve, corresponding to $t=1$,
is already close to a circle with radius $1/e^2\approx 0.14$.
%%%%%%%%%%%%%%%%%%%%%%%%%%%%%%%%%%%%%%%%%%%%%%%%%%%%%%%%%%%%
%% ELECTROSTATIC %%%%%%%%%%%%%%%%%%%%%%%%%%%%%%%%%%%%%%%%%%%%%%%%
%%%%%%%%%%%%%%%%%%%%%%%%%%%%%%%%%%%%%%%%%%%%%%%%%%%%%%%%%%%%
\section{The electrostatic model\label{sec:elec}}
%%%%%%%%%%%%%%%%%%%%%%%%%%%%%%%%%%%%%%%%%%%%%%%%%%%%%%%%%%%%
We now specialize the general equations~(\ref{uz})--(\ref{eq:sprop}) for the electrostatic equilibrium problem
to our case of interest, namely the case where $A=-1$ and the limit~(\ref{tdef}) exists.
As we have seen, the curves $\gamma_t$ for finite $t>0 $ are simple closed curves inside the  Szeg\H{o}
curve $\gamma_0$, and we have a $t$-dependent potential given by
 \begin{equation}
 	\label{utz}
	U_t(z) = \re z  + \log|z| + U_{t,{\log}}(z),
\end{equation} 
where the logarithmic potential of the density $\rho(z)$ is defined by
\begin{equation}
    \label{utlogz}
    U_{t,{\log}}(z) = -2 \int_{\gamma_t} \log |z-z'|\, \rho_t(z')  | d z'|.
\end{equation}
To compute $U_{t,{\rm log}}(z)$ we use equation~(\ref{ze40})  to write
\begin{equation}
	\label{plt1}
	\int_{\gamma_t}  \log |z-z'|\, \rho_t(z')| d z'| 
	=
	\re \int_{\gamma_t} \frac{ d z'}{2\pi i}  \log (z-z')\, \frac{1-z'}{z'} .
\end{equation} 
For $z\in D_t^-$ we take a branch of $\log (z-z')$ that is analytic for all $z'\in D_t^+$ and use
residues to find
\begin{equation}
	\label{plt3}
	\int_{\gamma_t} \frac{ d z'}{2\pi i}  \log (z-z') \frac{1-z'}{z'}
	=
	\log z,\quad z\in D_t^-. 
\end{equation} 
For $z\in D_t^+$ we follow an argument from Ref.~\cite{DI11}. The logarithmic potential is a continuous function
on $\mathbb{C}$ which is harmonic in $\mathbb{C}\setminus \gamma_t$. From equations~(\ref{utlogz}) and~(\ref{plt3})
we have that
\begin{equation}
    \label{plt40}
    U_{t,\log}(z) =- 2\log |z| ,\quad z\in D_t^-.
\end{equation} 
Hence,
\begin{equation}
    \label{plt4}
     U_{t,\log}(z) = -2\log |z| = -2\re z +2(t+1) ,\quad z\in \gamma_t,
\end{equation} 
and since  $U_{t,\log}(z) + 2 \re z $ is harmonic  in $ D_t^+$ we have that 
\begin{equation}
    \label{plt41}
    U_{t,\log}(z) = -2 \re z  + 2(t+1) ,\quad z\in D_t^+. 
\end{equation}
Thus, from equations~(\ref{plt40}) and~(\ref{plt41}) we obtain
\begin{equation}
	\label{plt8}
 	U_t(z) = \left\{\begin{array}{ll}
	                      \re z  - \log |z|,\quad \mbox{$z\in D_t^-$},  \\ \\
                             -\re z  + \log |z| + 2(t+1), \quad \mbox{$z\in D_t^+$}.
                         \end{array}\right.
\end{equation} 
The right and left limits of the potential function on $\gamma_t$ coincide
and are given by  $U_{t+}(z)=U_{t-}(z)= t+1$. Therefore~\cite{DI11},
\begin{equation}
\label{energy2}
 	u_0 = t+1.
\end{equation}
Note that from part~(b) of Theorem~1 we have that as $t\rightarrow +\infty$ the sources condensate
at $z=0$ and the total potential becomes
\begin{equation}
	\label{p0t}
	U_{\infty}(z) = \re z  - \log|z|.
\end{equation} 
Note also that because of equation~(\ref{la3}), the potential~(\ref{plt8})
is symmetric under the Schwarz reflection~(\ref{la1})
\begin{equation}
	\label{plt9}
	 U_t(z^*) = U_t(z),
\end{equation} 
which in fact is a consequence of the $S$ property. Note also that equations~(\ref{utlogz}) and~(\ref{plt40})
show explicitly that $\mu_\infty(z)$ is the electrostatic skeleton of $\mu_t(z)$.

Using equations~(\ref{plt4}) and~(\ref{plt41}) and the symmetry of $\gamma_t$ with respect to
the real axis, we can compute the self-energy $E_\mathrm{se}$ of the line conductor
\begin{equation}
    E_\mathrm{se}  =  \frac{1}{2}\int_{\gamma_t} U_{t,{\rm log}}(z)  \rho(z) |d z|
                              = t+1, \label{energy3dup}
\end{equation}
and using equations~(\ref{eu0mse}),  (\ref{energy2}) and~(\ref{energy3dup}) we  conclude that the total electrostatic
energy of the conductor vanishes,
\begin{equation}
	\label{energy}
	E = 0.
\end{equation}

\begin{figure}[h]
\centering
\includegraphics[width=0.75\textwidth]{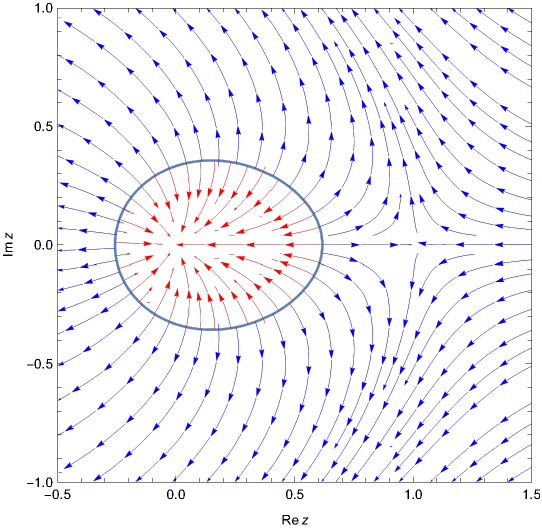}
\caption{Curve $\gamma_t$ corresponding to $t=1/10$ and electric field lines in $D_t^-$ and $D_t^+$
              according to Eq.~(\ref{ef1}).}\label{fig4}
\end{figure}

The corresponding electric field $\mathbf{E}_t = -\nabla U_t$  is given by  
\begin{equation}
      \label{ef1}
       \mathbf{E}_t(z)
       =
       \left\{\begin{array}{cc}
           \displaystyle -1+\frac{1}{\bar{z}}, \quad \mbox {$z\in D_t^-$},  \\ \\
           \displaystyle   1-\frac{1}{\bar{z}}, \quad \mbox{$z\in D_t^+$}. 
       \end{array}\right.
 \end{equation}
This electric field varies very quickly in a neighborhood of the corresponding curve $\gamma_t$, to the
extent that a vector plot is not very informative. Therefore, in Fig.~\ref{fig4} we show the field lines and
the curve $\gamma_t$ corresponding to $t=1/10$. Note the vanishing electric field at $z=1$.
 
\begin{figure}[h]
\centering
\includegraphics[width=0.75\textwidth]{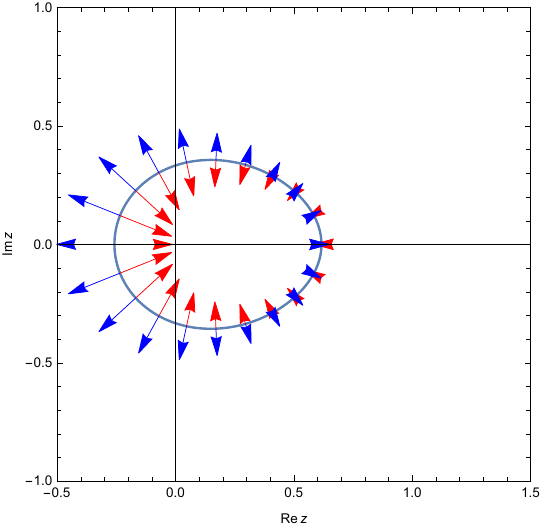}
\caption{Curve $\gamma_t$ corresponding to $t=1/10$ and electric field $\mathbf{E}_t(z) $ scaled down
by a factor of 20 with respect to the marks on the axes, illustrating how the electrostatic force acting on the points
of $\gamma_t$ vanishes, i.e., how the $S$ property of of Stahl~\cite{ST85a,ST85b} and Gonchar and Rakhmanov~\cite{GO89}
is satisfied.}\label{fig5}
\end{figure}

 In an attempt to illustrate the $S$ property~(\ref{eq:sprop}) of Stahl~\cite{ST85a,ST85b} and
 Gonchar and Rakhmanov~\cite{GO89}, in Fig.~\ref{fig5} we show the electric field on the curve $\gamma_t$
 scaled down by a factor of $20$ with respect to the marks on the axes. Indeed,  for $z\in\gamma_t$
 the right and left limits of the electric field verify $\mathbf{E}_t^{(-)}(z)=-\mathbf{E}_t^{(+)}(z)$
 and therefore the electrostatic force acting on the points of $\gamma_t$ vanishes.

%%%%%%%%%%%%%%%%%%%%%%%%%%%%%%%%%%%%%%%%%%%%%%%%%%%%%%%%%%%%
%%%%%%%%%%%%%%%%%%%%%%%%%%%%%%%%%%%%%%%%%%%%%%%%%%%%%%%%%%%%

%%%%%%%%%%%%%%%%%%%%%%%%%%%%%%%%%%%%%%%%%%%%%%%%%%%%%%%%%%%%
%%%%%%%%%%%%%%%%%%%%%%%%%%%%%%%%%%%%%%%%%%%%%%%%%%%%%%%%%%%%

%%%%%%%%%%%%%%%%%%%%%%%%%%%%%%%%%%%%%%%%%%%%%%%%%%%%%%%%%%%%
\subsection{Conformal transformation onto the $\mathcal{R}_w$ Riemann surface\label{sec:rs}}
%%%%%%%%%%%%%%%%%%%%%%%%%%%%%%%%%%%%%%%%%%%%%%%%%%%%%%%%%%%%
We may now perform a conformal transformation of the electrostatic model from the $z$-plane
onto the Riemann surface $\mathcal{R}_w$. The complex potential corresponding to (\ref{plt8}) is
\begin{equation}
    \label{cp}
    \Omega_t(z) = \left\{\begin{array}{ll}
                                        z-\log z, \quad \mbox{$z\in D_t^-$},  \\ \\
                                       -z+\log z+2(t+1), \quad \mbox{$z\in D_t^+$},
                            \end{array}\right.
\end{equation}
and the branches of the transformed complex potentials $\hat{\Omega}_{t,k}(w)$ on the cut $w$-planes
$\mathbb{C}_{w,k}\equiv\mathbb{C}_w\setminus \omega_k$ are
\begin{equation}
    \label{ct}
    \hat{\Omega}_{t,0}(w) = \left\{\begin{array}{ll}
                                             -\log w+1, \quad |w|>e^{-t}, \\ \\
                                              \log w+2t+1, |w|<e^{-t},
                                   \end{array}\right.
 \end{equation} 
and 
\begin{equation}
    \label{ct0}
     \hat{\Omega}_{t,k}(w) =-\log w+1, \quad k\in\mathbb{Z}\setminus\{0\}.
 \end{equation} 
 The corresponding electrostatic potentials $\hat{U}_{t,k}(w) = \re \hat{\Omega}_{t,k}(w)$ are
 \begin{equation}
     \label{ctr}
     \hat{U}_{t,0}(w) = \left\{\begin{array}{ll}
                                      -\log |w|+1, \quad |w|>e^{-t},  \\ \\
                                       \log |w|+2t+1, \quad|w|<e^{-t},
                          \end{array}\right.
 \end{equation} 
and 
\begin{equation}
    \label{ctro}
     \hat{U}_{t,k}(w) = -\log |w|+1, \quad k\in\mathbb{Z}\setminus\{0\}.
\end{equation} 
 
The density~(\ref{ze40}) transforms as~\cite{DI11}
\begin{equation}
    \label{ze401}
    \rho_t(z)|{ d} z| = \frac{1}{2\pi i}\frac{1-z}{z}{ d} z = \frac{1}{2 \pi i}\frac{ d w}{w},
\end{equation}
which represents a uniform unit charge density on the circle $|w|=e^{-t}$, and gives rise to a logarithmic potential
\begin{equation}
    \label{ctl}
    \hat{U}_{t,\log}(w) = \left\{\begin{array}{ll}
                                       -2\log |w|, \quad |w|>e^{-t},  \\ \\
                                        2 t, \quad |w|<e^{-t}.
                                    \end{array}\right.
 \end{equation} 
 Therefore, in view of equations~(\ref{ctr}), (\ref{ctro}) and~(\ref{ctl}) the conformal image of the model
 on the sheet $\mathbb{C}_{w,0}$ is the superposition of the potential $\log| w|+1$ due to a point charge
 $q=-1/2$ at $w=0$ and the potential created by a unit charge uniformly distributed on the circle $|w|=e^{-t}$.
 On the remaining sheets $\mathbb{C}_{w,k}$, $k\in\mathbb{Z}\setminus\{0\}$, the model represents
 the potential of a point charge $q=1/2$ at  $w=0$.

We notice that  the background external field $\re z $ of the model in the $z$-plane disappears in the transformed
model, which is radially symmetric in the Riemann surface $\mathcal{R}_w$.  In particular, the Schwarz
reflection symmetry  of the model in the $z$-plane becomes the symmetry under inversion
with respect to the circle $|w|=e^{-t}$ in the sheet $\mathbb{C}_{w,0}$.
%%%%%%%%%%%%%%%%%%%%%%%%%%%%%%%%%%%%%%%%%%%%%%%%%%%%%%%%%%%%
\subsection{Dual hydrodynamical model}
%%%%%%%%%%%%%%%%%%%%%%%%%%%%%%%%%%%%%%%%%%%%%%%%%%%%%%%%%%%%
It is interesting to consider briefly the dual hydrodynamical model determined by the complex potential $i \Omega_t(z)$.
The  velocity field is defined by
\begin{equation}
   \label{vt}
   \everymath{\displaystyle}
    \mathbf{v}_t(z)
    =
    -\nabla \im \Omega_t(z)
    =
    \left\{
    \begin{array}{c}
        -i\left(1-\frac{1}{\bar{z}}\right)
        =
        -i\frac{|z|^2-z}{|z|^2}, \quad z \in D^-_t, \\
        i\left(1-\frac{1}{\bar{z}}\right)
        =
        i\frac{|z|^2-z}{|z|^2}, \quad z \in D^+_t,
    \end{array}
    \right.
\end{equation}
and we have a vortex density on $\gamma_t$. Since the tangent vector to $\gamma_t$ is given by,
\begin{equation}
    z_{\theta} = -i\frac{z}{z-1} = -i\frac{|z|^2-z}{|z-1|^2},
\end{equation}
the $S$-property of $\gamma_t$ implies that the right and left
limits of the velocity field are tangent to $\gamma_t$ and satisfy $\mathbf{v}_t^{(-)}(z)=-\mathbf{v}_t^{(+)}(z)$.
The model describes a fluid flowing outside and inside a hollow obstacle  represented by $\gamma_t$,
and since the pressure $P_t(z)$ at each point $z$ is proportional to $|\mathbf{v}_t(z)|^2$,
the $S$-property implies  that the net force per unit length acting on a point of  $\gamma_t$ vanishes.

\begin{figure}[ht]
\centering
\includegraphics[width=0.9\textwidth]{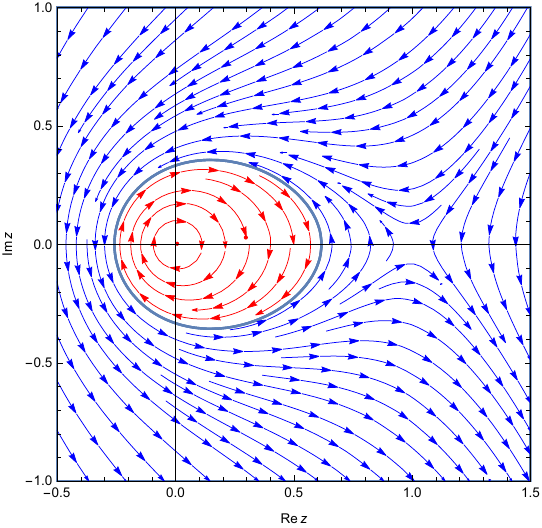}
\caption{Streamlines corresponding to the velocity vector field $\mathbf{v}_t$ of Eq.~(\ref{vt}) for
              $t=1/10$. Note the stagnation point at $z=1$.}
\label{fig6}
\end{figure}

Again, the velocity vector field varies too quickly on a neighborhood of the curve to permit an
illustrative picture. Therefore, in Fig.~\ref{fig6} we show the streamlines corresponding
to $t=1/10$.

%%%%%%%%%%%%%%%%%%%%%%%%%%%%%%%%%%%%%%%%%%%%%%%%%%%%%%%%%%%%
%%%%%%%%%%%%%%%%%%%%%%%%%%%%%%%%%%%%%%%%%%%%%%%%%%%%%%%%%%%%

%%%%%%%%%%%%%%%%%%%%%%%%%%%%%%%%%%%%%%%%%%%%%%%%%%%%%%%%%%%%
 \section{The Penner matrix model\label{sec:mm}}
%%%%%%%%%%%%%%%%%%%%%%%%%%%%%%%%%%%%%%%%%%%%%%%%%%%%%%%%%%%%
%%%%%%%%%%%%%%%%%%%%%%%%%%%%%%%%%%%%%%%%%%%%%%%%%%%%%%%%%%%%
 \subsection{The saddle point equations for a random matrix model}
%%%%%%%%%%%%%%%%%%%%%%%%%%%%%%%%%%%%%%%%%%%%%%%%%%%%%%%%%%%%
The partition function~(\ref{mm}) can be rewritten as
\begin{equation}
	\label{mmm}
	Z_{n}(g_n)
	=
	\frac{1}{n!}	\int_{\Gamma \times\cdots\times\Gamma} \exp\big(-n^2\mathcal{S}_n(\mathbf{z})\big)
	\prod_{i=1}^n d z_i,
\end{equation}
where $\mathbf{z}=(z_1,\dots,z_n)$ and
\begin{equation}
	\label{inte}
	\mathcal{S}_n(\mathbf{z}) = \frac{1}{g_n n^2}\sum_{i=1}^n W(z_i) - \frac{1}{2 n^2}\sum_{i=1}^n\sum_{ j\neq i}\log(z_i-z_j)^2,
\end{equation}
and the corresponding saddle points are the solutions of the equations
\begin{equation}
	\label{sad}
	\frac{\partial\mathcal{S}_n}{\partial z_i}(\mathbf{z}) = 0,	
	\quad
	i=1,\ldots, n,
\end{equation}
or explicitly
\begin{equation}
	\label{sa}
	\frac{1}{g_n}W'(z_i)+\sum_{j\neq i}\frac{2}{z_j-z_i}=0,
	\quad
	i=1,\ldots,n.
\end{equation}
Note that these equations are symmetric under permutations of the $z_i$,
and therefore generically a solution gives rise to a set of $n!$ solutions obtained by
permutations.

As a consequence of~(\ref{sa}), the resolvent functions $\omega_n(z)$ for the sequence of monic polynomials~(\ref{fep})
\begin{equation}
	\label{re}
	\omega_n(z) = \frac{1}{n}\frac{S'_n(z)}{S_n(z)}
	                     = \frac{1}{n}\sum_{i=1}^n \frac{1}{z-z_i^{(n)}}
\end{equation} 
satisfies the Riccati equation
\begin{equation}
	\label{ri}
	\frac{1}{n}\omega'_n(z) + \omega_n(z)^2-\frac{1}{n g_n}W'(z)\omega_n(z)
	=
	-\frac{1}{n^2 g_n}\sum_{i=1}^n\frac{W'(z)-W'(z_i^{(n)})}{z-z_i^{(n)}}.
\end{equation}
%%%%%%%%%%%%%%%%%%%%%%%%%%%%%%%%%%%%%%%%%%%%%%%%%%%%%%%%%%%%
 \subsection{Zero counting measures for the associated monic polynomials and the $n\to\infty$ limit}
%%%%%%%%%%%%%%%%%%%%%%%%%%%%%%%%%%%%%%%%%%%%%%%%%%%%%%%%%%%%
The saddle point method assumes the existence of a sequence of saddle points
\begin{equation}
    \big\{{\mathbf{z}}^{(n)}=(z_1^{(n)},\ldots,z_n^{(n)})\big\}_{n=1}^\infty
\end{equation}
of  $\mathcal{S}_n$ which in turn determine the sequence of monic polynomials~(\ref{fep})
and zero counting measures~(\ref{me1}).
Note that except for the factor $1/n$, the resolvent function~(\ref{re}) is the Stieltjes (Cauchy) transform of $\nu_n$
\begin{equation}
   \label{Cdis}
   \omega_n(z) = \frac{1}{n} \int_{\mathbb{C}}\frac{d\nu_n(z')}{z-z'}.
\end{equation}    
Therefore, if all the saddles $\mathbf{z}^{(n)}$ belong to the same compact set $K\subset\mathbb{C}$,
and taking into account~(\ref{re}) and~(\ref{Cdis}), we have that
\begin{equation}
  \label{omeganlim}
  \omega(z) = \lim_{n\in\Delta}\omega_n(z)
                   = \int_{\mathbb C} \frac{d\mu(z')}{z-z'},\quad z\in\mathbb{C}\setminus\operatorname{supp}(\mu).
\end{equation}
%%%%%%%%%%%%%%%%%%%%%%%%%%%%%%%%%%%%%%%%%%%%%%%%%%%%%%%%%%%%
 \subsection{The Schwinger-Dyson equation}
%%%%%%%%%%%%%%%%%%%%%%%%%%%%%%%%%%%%%%%%%%%%%%%%%%%%%%%%%%%%
The saddle point method assumes that there exists a unit-normalized positive density
$\rho_{\text{MM}}(z)$ with support $\hat{\gamma}$ such that
\begin{equation}
   \label{murho}
   d\mu(z) = \rho_{\text{MM}}(z)|dz|,
\end{equation}
and consequently the weak-* limit $\omega(z)$ of the resolvent $\omega_n(z)$ is the Cauchy transform of the measure
\begin{equation}
	\label{ef}
	\omega(z) = \int_{\hat{\gamma}}\frac{\rho_{\text{MM}}(z') |d z'|}{z-z'}.
\end{equation}
Therefore, as a consequence of~(\ref{tH}) and~(\ref{ri}), the function $\omega(z)$ must satisfy the Schwinger-Dyson equation
\begin{equation}
 	\label{rl}
	\omega(z)^2-\frac{1}{T}W'(z)\omega(z)
	=
	-\frac{1}{T}\int_{\hat{\gamma}}\frac{W'(z)-W'(z')}{z-z'}\rho_{\text{MM}}(z') |d z'|.
\end{equation}

If we define
\begin{equation}
	\label{lai0}
	y_{\text{MM}}(z) = \frac{1}{T} W'(z) - 2 \omega(z),
\end{equation}
\begin{equation}
	\label{ere0}
	R_{\text{MM}}(z) = \Big(\frac{1}{T}W'(z)\Big)^2-\frac{4}{T} \int_{\hat{\gamma}}\frac{W'(z)-W'(z')}{z-z'}\rho_{\text{MM}}(z') |d z'|,
\end{equation}
the Schwinger-Dyson equation~(\ref{rl}) can be rewritten as
\begin{equation}
	\label{lay}
	y_{\text{MM}}(z)^2 = R_{\text{MM}}(z).
\end{equation}
Three comments are in order: (i) under our assumptions on $W'(z)$, the function $R_{\text{MM}}(z)$ defined in~(\ref{ere0})
is a rational function of $z$ with poles at the same points as those of $W'(z)$;
(ii) the density $\rho_{\text{MM}}(z)$ can be recovered from $y_{\text{MM}}(z)$ using the Sokhotskii-Plemelj formulas;
and (iii) the function $y_{\text{MM}}(z)$ is analytic outside $\hat{\gamma} \cup \mathcal{A}$, and~(\ref{lay}) implies that
\begin{equation}
	\label{rh}
	y_{\text{MM}}(z_+) = -y_{\text{MM}}(z_-),\quad z\in \hat{\gamma}.
\end{equation}

Moreover, from equations~(\ref{lai0})--(\ref{lay}) and~(\ref{ef}) it follows that
 \begin{equation}
 	\label{id}
 	R_{\text{MM}}(z) = \Big(\frac{1}{T} W'(z) - 2 \int_{\hat{\gamma}}\frac{\rho_{\text{MM}}(z') |d z'|}{z-z'}\Big)^2,
\end{equation}
which shows that $\rho_{\text{MM}}(z)|dz|$ is a continuous critical measure on $\mathbb{C}$
in the sense of Mart\'{\i}nez-Finkelshtein and Rakhmanov~\cite{MA11}.  As a consequence
(see Lemma~5.2 of~\cite{MA11} and Proposition~3.8 of~\cite{KU14}), the support $\hat{\gamma}$ of
$\rho_{\text{MM}}(z)$ is a  union of a finite number of analytic arcs
\begin{equation}
	\label{cuts}
	\hat{\gamma} = \hat{\gamma}_1\cup\hat{\gamma}_2\cup \cdots\cup \hat{\gamma}_s,
\end{equation}
which are maximal trajectories of the quadratic differential 
\begin{equation}
	\label{qa}
	 -R_{\text{MM}}(z)(d z)^2,
\end{equation} 
i.e., maximal curves~\cite{ST84} $z=z(\tau)\, (\tau\in (\alpha,\beta))$ such that
\begin{equation}
	-R_{\text{MM}}(z) \Big(\frac{d z}{d \tau}\Big)^2>0, \quad \mbox{for all $\tau\in (\alpha,\beta)$}.
\end{equation}
%%%%%%%%%%%%%%%%%%%%%%%%%%%%%%%%%%%%%%%%%%%%%%%%%%%%%%%%%%%%
 \subsection{Particularization to the Penner matrix model}
%%%%%%%%%%%%%%%%%%%%%%%%%%%%%%%%%%%%%%%%%%%%%%%%%%%%%%%%%%%%
Let us particularize the results of the previous section for the Penner matrix model defined by the potential~(\ref{eq:pmm})
with {}'t~Hooft parameter $T$.

Equation~(\ref{ere0}) takes the form
\begin{equation}
	\label{ere0P}
	R_\text{PM}(z) = \frac{1}{T^2}\left(1+\frac{1}{z}\right)^2
	                          + \frac{4}{Tz}\int_{\hat{\gamma}}\frac{1}{z'}\rho_\text{PM}(z') |dz'|.
\end{equation}
To determine $R_\text{PM}(z)$ note that
\begin{equation}
   \label{Casy}
   \int_{\hat{\gamma}}\frac{\rho_\text{PM}(z') |d z'|}{z-z'} = \frac{1}{z} + \mathcal{O}\left(\frac{1}{z^2}\right)
   \quad\mbox{as } z\rightarrow\infty,
\end{equation}
and therefore
\begin{equation}
 	\label{idP}
 	\begin{aligned}
 	R_\text{PM}(z)  = \frac{1}{T^2}\left(1+\frac{1}{z}\right)^2-\frac{4}{Tz}+\mathcal{O}\left(\frac{1}{z^2}\right)
	\quad \mbox{as }
 	   z\rightarrow\infty.
 	\end{aligned}
\end{equation}
From equations~(\ref{ere0P}) and~(\ref{idP}) it follows that
\begin{equation}
\int_{\hat{\gamma}}\frac{1}{z'}\rho_\text{PM}(z') |dz'| = -1,
\end{equation}
and that
\begin{equation}
   \label{RPenner}
    R_\text{PM}(z)=\frac{1}{T^2}\left(1+\frac{1}{z}\right)^2-\frac{4}{Tz}.
\end{equation}
In particular, for the critical case $T=1$ we recover the result~(\ref{eq:rcm})
\begin{equation}
   \label{RPennerT1}
    R_\text{PM}^\text{crit}(z)=\left(1-\frac{1}{z}\right)^2.
\end{equation}
%%%%%%%%%%%%%%%%%%%%%%%%%%%%%%%%%%%%%%%%%%%%%%%%%%%%%%%%%%%%
\subsection{The Penner matrix model and the Laguerre polynomials}
%%%%%%%%%%%%%%%%%%%%%%%%%%%%%%%%%%%%%%%%%%%%%%%%%%%%%%%%%%%%
The saddle point equations~(\ref{sa}) for the Penner model are
\begin{equation}
	\label{sa0}
	\frac{1}{g_n}\Big(1+\frac{1}{z_i^{(n)}}\Big)+\sum_{j\neq i}\frac{2}{z_j^{(n)}-z_i^{(n)}}=0,\quad i=1,\ldots,n,
\end{equation}
and the corresponding Riccati equation~(\ref{ri}) is 
\begin{equation}
	\label{ripl}
	\frac{1}{n}\omega'_n(z)+\omega_n(z)^2-\frac{1}{n g_n}\Big(1+\frac{1}{z} \Big)\omega_n(z)
	=
	\frac{1}{n^2 g_n\,z}\sum_{i=1}^n \frac{1}{z_i^{(n)}}.
\end{equation}
From~(\ref{sa0}) it follows that
\begin{equation}
	\label{idpl}
	\sum_{i=1}^n \frac{1}{z_i^{(n)}}=-n,
\end{equation}
and we get the following  second order linear equation for $S_n(z)$
\begin{equation}
	\label{ri1b}
	S''_n(z)-\frac{1}{ g_n} \Big(1+\frac{1}{z} \Big)S'_n(z) = -\frac{n}{g_n\,z}S_n(z).
\end{equation}
By comparing~(\ref{ri1b}) with  the Laguerre differential equation
\begin{equation}
	\label{lage}
	z u''(z)+(\alpha+1-z) u'(z)+n u(z)=0,
\end{equation}
we find that the monic polynomials $S_n(z)$ are proportional to the rescaled Laguerre polynomials
$L^{(\alpha_n)}_n(z/g_n)$ where $\alpha_n = - 1 - 1/g_n$. Therefore, the saddle points
${\mathbf{z}}^{(n)}=(z_1^{(n)},\ldots,z_n^{(n)})$ of the Penner model with coupling constants
$g_n$ are given by
\begin{equation}
	\label{sopl}
	z_i^{(n)} = g_n\,l^{(\alpha_n,n)}_i,\quad i=1,\ldots,n,
\end{equation}
where  $l^{(\alpha,n)}_i$ are  the zeros of $L^{(\alpha)}_n(z)$.
%%%%%%%%%%%%%%%%%%%%%%%%%%%%%%%%%%%%%%%%%%%%%%%%%%%%%%%%%%%%q
 \subsection{The {}'t~Hooft limit of the critical Penner matrix model}
%%%%%%%%%%%%%%%%%%%%%%%%%%%%%%%%%%%%%%%%%%%%%%%%%%%%%%%%%%%%
The large $n$ limit of the Laguerre polynomials is related to the {}'t~Hooft limit of the Penner matrix model~\cite{AL15}
under the identifications
\begin{equation}
    \label{ide}
    \alpha_n = -1 - \frac{1}{g_n},\quad A=-\frac{1}{T}.
\end{equation}
Therefore, the eigenvalue density $\rho_\text{PM}(z)$ of the Penner matrix model and the zero distribution
$\rho_t(z)$ of the scaled Laguerre polynomials $L^{(\alpha_n)}_n(nz)$ are related by
\begin{equation}
   \rho_\text{PM}(z) = \frac{1}{|T|}\rho_t\left(\frac{z}{T}\right),
\end{equation}
and in particular the large $n$ limit of the Laguerre polynomials with $A=-1$ corresponds to the {}'t~Hooft
limit of the Penner matrix model with $T=1$, which in turn describes the critical case of the large $n$
Penner model~\cite{PA10}. 
%%%%%%%%%%%%%%%%%%%%%%%%%%%%%%%%%%%%%%%%%%%%%%%%%%%%%%%%%%%%
 \section{Summary}
%%%%%%%%%%%%%%%%%%%%%%%%%%%%%%%%%%%%%%%%%%%%%%%%%%%%%%%%%%%%
We have analyzed the one-parameter family of deformations of the classical Szeg\H{o} curve $\gamma_0$ given by
$\gamma_t  = \{ z\in\mathbb{C}: |z\, e^{1-z}| = e^{-t}, |z|\leq 1\}$, $t\geq 0$ from three different viewpoints:
as supports of equilibrium measures in an external electrostatic field, as the dual hydrodynamic model,
and as supports of the limiting zero counting measures of certain subsequences of Laguerre polynomials,
which appear in particular as limiting supports of the saddle points in the critical case of the Penner matrix model.

We discuss the shrinking process $t\to\infty$ using as our main tool the Schwarz $S$ functions of the curves $\gamma_t$.
In general, $S$ functions are not available in closed form and their domains of analyticity are difficult to determine
(apart from the standard fact that  $S$ is analytic in a neighborhood of the curve). In our setting, however,
the Schwarz function can be expressed explicitly in terms of the Lambert  $W$ function, and its domain of analyticity
can likewise be described in explicit terms. Moreover, in this formulation the  $S$-property of Stahl~\cite{ST86a,ST86b}
and Gonchar and Rakhmanov~\cite{GO89}, which essentially governs the determination of the support,
can be written in explicit form as the Schwarz reflection symmetry.

In particular, the potential functions, electric fields, and characteristic energies in the electrostatic formulation
can all be computed explicitly, as can the complex potential and velocity field in the dual hydrodynamical description, and 
in both cases the $S$-property acquires the natural physical interpretations: in the electrostatic formulation,
that the net electrostatic force on the conductor $\gamma_t$ vanishes, and in the hydrodynamic interpretation,
 that the net force per unit length acting on any point of the curve $\gamma_t$ vanishes.
%%%%%%%%%%%%%%%%%%%%%%%%%%%%%%%%%%%%%%%%%%%%%%%%%%%%%%%%%%%%%%%%%%%%%%%
\backmatter
%%%%%%%%%%%%%%%%%%%%%%%%%%%%%%%%%%%%%%%%%%%%%%%%%%%%%%%%%%%%%%%%%%%%%%%
\section*{Acknowledgements}
This work was partially supported by grants PID2024-155527NB-I00 from Spain's Ministerio de
Ciencia, Innovaci\'on y Universidades and~PR12/24-31565 from Universidad Complutense de Madrid.

We thank Prof.~A.~Mart\'{\i}nez  Finkelshtein for useful conversations and for calling
our attention to the available results on zeros of Laguerre polynomials.
\%% BioMed_Central_Bib_Style_v1.01

%%%%%%%%%%%%%%%%%%%%%%%%%%%%%%%%%%%%%%%%%%%%%%%%%%%%%%%%%%%%%%%%%%%%%%%
\end{document}